\documentclass[manuscript]{aastex63}

\received{December 31, 2019}
\accepted{February 7, 2020}

\submitjournal{Astronomical Journal}

\shorttitle{Cloud Identification with Machine Learning}
\shortauthors{Mommert}

\begin{document}

\title{Cloud Identification from All-sky Camera Data with Machine Learning}

\correspondingauthor{Michael Mommert}
\email{michael.mommert@lowell.edu}

\author[0000-0002-0786-7307]{Michael Mommert}
\affiliation{Lowell Observatory \\
1400 W Mars Hill Rd, Flagstaff AZ, 86001, USA}

\begin{abstract}
  Most ground-based observatories are equipped with wide-angle all-sky
  cameras to monitor the night sky conditions. Such
  camera systems can be used to provide early warning of incoming clouds that
  can pose a danger to the telescope equipment through precipitation, as well
  as for sky quality monitoring.
  We investigate the use of different machine learning approaches for
  automating the identification of mostly opaque clouds in all-sky camera
  data as a cloud warning system.  In
  a deep-learning approach, we train a Residual Neural Network (ResNet)
  on pre-labeled camera images. Our second approach extracts relevant
  and localized image features from camera images and uses these data
  to train a gradient-boosted tree-based model (lightGBM). We train
  both model approaches on a set of roughly 2,000 images taken by the all-sky
  camera located at Lowell Observatory's Discovery Channel Telescope,
  in which the presence of clouds has been labeled manually.
  The ResNet approach reaches an
  accuracy of 85\% in detecting clouds in a given region of an
  image, but requires a significant amount of computing resources.
  Our lightGBM approach achieves an accuracy of 95\%
  with a training sample of ${\sim}$1,000 images and rather modest
  computing resources. Based on different performance metrics, we
  recommend the latter feature-based
  approach for automated cloud detection. Code
  that was built for this work is available online.
\end{abstract}


\section{Introduction} 
\label{sec:introduction}

Ground-based telescopes are exposed to and have to be protected from
environmental influences. Precipitation and high relative humidity can
cause significant damage to optical surfaces, the telescope structure,
as well as telescope electronics. In order to protect telescopes from
rain and snow, the most conservative policy is to close their
enclosures as soon as the sky is clouded out.

All-sky cameras provide an efficient and inexpensive means to monitor
cloud coverage at night. Such cameras, which often use inexpensive
charge-coupled-device (CCD) or complementary metal–oxide–semiconductor
(CMOS) detectors in combination with wide-angle lenses provide the
sensivity and dynamic range that is necessary to identify cloud
coverage even on a dark night. Most observatories -- ranging from
small aperture telescopes to the largest available telescopes --
already have such camera systems installed and use them on a regular
basis.

Typically, human telescope operators monitor all-sky camera feeds in
real-time, allowing them to react to incoming clouds by closing the
telescope enclosure on short time scales. While the task of
identifying clouds against the sky background is mostly trivial for
humans, this is a non-trivial task for a machine. Problems arise
because of the variable appearances clouds can have during the
night. Depending on illumination conditions from the Moon or the Sun,
clouds can either appear brighter than the clear sky, or
darker. Furthermore, at the low imaging resolutions provided by all-sky
cameras even the clear sky itself contains bright and dark patches,
depending on the density and brightness of stars in a given field, creating a source of confusion in cloud
identification. Additional complications arise from static or variable
effects from terrestrial illumination and the observatory's local
horizon. A simple classification scheme, for instance based on sky
brightness, is in most cases not sufficient to identify clouds with
high confidence.

While the main motivation for this work is the development of a system
that provides warning in the presence of opaque clouds that can potentially
carry precipitation, the same
methodology can be applied to quantify sky quality. Such a
system can be used to derive the fraction of the accessible sky that is clear
throughout a night or longer periods, or to identify near-photometric
conditions without human interaction.

The automation of cloud discovery from all-sky camera data is
worthwhile due to the ubiquity and general availability of these
camera systems. An automated system that is able to warn a robotic
telescope -- or a human operator -- of incoming clouds will
significantly improve the safety of observatories and enable automated
monitoring of sky quality.

In this work, we investigate the use of machine learning methods to
identify and locate clouds in all-sky camera data using two different
approaches. In our first approach, we use a deep-learning approach based on a
residual neural network
(ResNet) model that works on image data as obtained from the
camera. Our second approach combines the extraction of carefully
designed features that are indicative of clouds from images with a
tree-based machine learning model (lightGBM). 

Our requirement is to reach an accuracy of 95\% for the identification
of clouds for a given location on the sky from all-sky camera image
data.

We define our models in Section \ref{sec:models}, present their
results in Section \ref{sec:results}, and discuss their performances
and possible application to different use cases in Section
\ref{sec:discussion}.
Appendix \ref{sec:implementation}
briefly discusses the implementation of the models and the code used
in this work.

\section{All-sky Camera Data}

\label{sec:data}

\begin{figure}[t]
 \centering
 \gridline{\fig{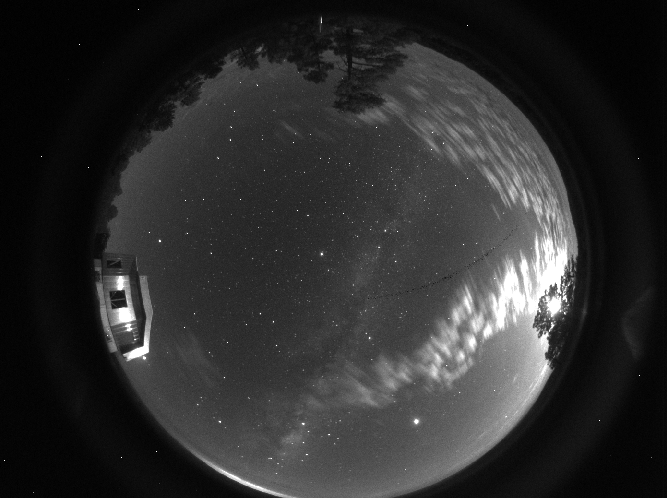}{0.4\textwidth}{(a) Raw Image}
           \fig{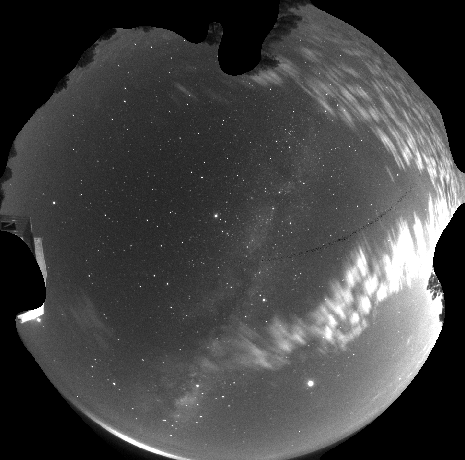}{0.3\textwidth}{(b) Cropped and Masked Image}}
 \gridline{\fig{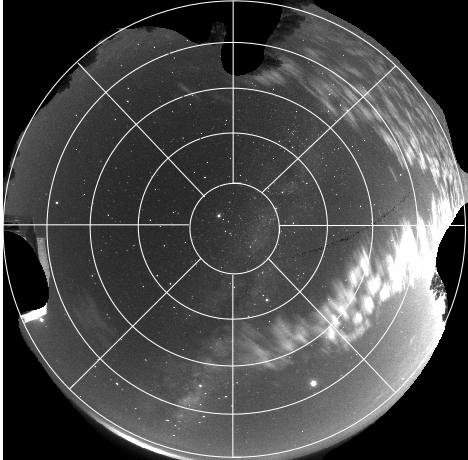}{0.3\textwidth}{(c) Subregion Grid}
           \fig{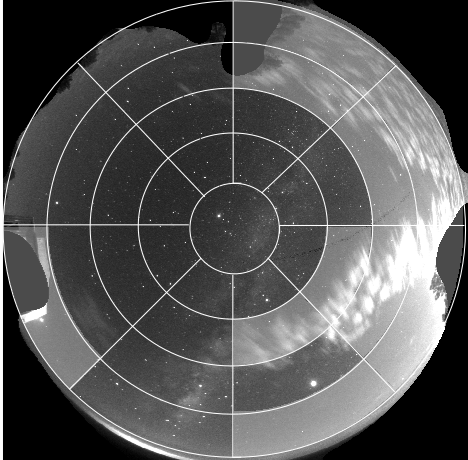}{0.3\textwidth}{(d) Labeled Subregions}}
 \caption{Preparation stages for our image data. (a) Raw image showing the circular projection of the sky and illumination on lens assembly parts. Clouds appear bright due to the imminent rise of the Moon; the Milkyway is visible, too. (b) Cropped image in which non-relevant parts for our task have been masked (telescope enclosure, nearby trees, and lens assembly). (c) Subregion grid resulting in 33 subregions. (d) Labeled subregion grid (subregions that contain significant amounts of clouds are highlighted).}
\label{fig:example_image}
\end{figure}

We train and test our machine learning models on image data taken at
Lowell Observatory's Discovery Channel Telescope. The images have been
obtained with a Starlight Xpress Oculus all-sky camera, featuring a
1392$\times$1040 pixel CCD detector and a 1.55~mm
f/1.2 fish-eye lens with a field of view of 180\degr. The camera
creates a circular projection of the sky plane with a radius of
520~pixels and 16-bit dynamic range. Image files are provided in
FITS format; the image header includes information on the date and
time of the observation and the exposure time (typically 60~s at
night). The imaging cadence at night is 1~min$^{-1}$. Figure
\ref{fig:example_image} (panel a) shows a raw example image from this
camera.

\subsection{Data Preparation}
\label{sec:data_preparation}

From each image we crop a quadratic region that contains the
circular image of the sky. 

We mask those parts of the image that do
not show the sky -- including parts of the lens assembly, as well as
local background features like the telescope enclosure and trees -- by
settting the corresponding pixel values to zero. 
The mask is created from a median combination of 50 random images that
were taken under similar (bright) illumination conditions. We blur the
resulting combined image with a Gaussian filter ({\tt
  scipy.ndimage.filters.gaussian\_filter}) to remove small-scale
features and use a threshold to extract those parts of the image that
do not show the sky. Finally, we smooth the edges of our selection by
convolving the resulting mask with a square kernel. The resulting mask
has been applied in Figure \ref{fig:example_image} (panel b). While
this approach does not perfectly mask all objects on the local
horizon, it is sufficient for our purposes.

To roughly localize clouds in the image data, we divide each image
into a set of subregions. The borders of these subregions are defined
in terms of radial distance from Zenith and azimuth. The innermost
subregion is a circular aperture centered on Zenith; more distant
areas are arranged in rings that are sub-divided into equidistant
ring-segments based on azimuth. This radially symmetric definition has
the advantage that ring segments on the same ring correspond to the
same elevation and the same airmass. Furthermore, this definition
naturally reflects the ranked importance of finding clouds at
different elevations: while clouds close to Zenith may pose an
immediate threat to the observatory, clouds on the horizons are no
immediate threat but should be recognized and monitored. This ranking can be
directly translated into different warning levels.

For our data, we chose a scheme that consists of a circular aperture around
Zenith, 3 rings, and 8 ring-segments. The distribution of the 33 resulting
subregions is shown in Figure \ref{fig:example_image} (panel c).

\subsection{Training Data Sample}
\label{sec:data_training}

Our training data sample consists of 1,975 images that were randomly
drawn from a set of 259,259 images taken between June 2018 and August
2019. For each image we manually label the presence of clouds in each
subregion with a binary flag. Subregions are assigned unity value if
they contain considerable amounts
of clouds that significantly affect the transparency in the
corresponding subregions (see Section \ref{sec:cloud_definition} for a
discussion), or zero otherwise; thick cirrus affecting transparency is
considered as a
cloud, while thin cirrus maybe not be considered as a cloud.  Figure
\ref{fig:example_image} (panel d) shows a labeled example image.
 
From our 1,975 training data images, we extract 65,175 labeled
subregions, 28,872 (44.3\%) of which contain clouds, 36,303 (55.7\%)
of which do not contain clouds. Hence, our training data sample has a
slight class imbalance that favors clear sky conditions.

\section{Machine Learning Model Definitions}
\label{sec:models}

We use two different approaches and two different machine learning
models to learn and predict the presence of clouds in our image data:
a residual neural network (ResNet) that works with image data, and a
tree-based model (lightGBM) that works on a set of carefully designed
features that we extract from each image and subregion.

\subsection{Image-Based Approach (ResNet)}

We use a residual neural network \citep[ResNet,][]{Kaimang2015}
adaptation that works directly on the cropped and masked image
data. ResNets are frequently used in computer vision applications,
including object identification, localization, classfication, and
image segmentation. 

Mimicking the way in which biological neural
networks (e.g., the human brain) work, artificial neural networks
consist of several layers of "neurons" that are connected with each
other and react to external stimulation in the form of an input data
vector. Each neuron is a mathematical function that uses a weighting
scheme to calculate a scalar output value based on the weights, the
input data vector, and a non-linear activation function.  In simple
feed-forward neural networks, each neuron is fully connected to all neurons in
the previous layer and all neurons in the following layer; outputs of the
previous layer serve as input for the current layer, while the output
of the current layer serves as input for the following layer. By
carefully designing a neural network and training it with ground-truth
data, it can learn patterns and perform tasks like image
classification. The learning process is an optimization process that
adjusts the weights in each neuron such that the output of the model
agrees with the ground-truth. See \citet{Russel2009} and
\citet{Goodfellow2016} for a review of the details of neural networks.

ResNets have a more complex network architecture including a number of
convolutional layers and the outputs of neurons do not only
affect the neurons in the following layer, but also those in later
layers. This principle enables the training of extremely deep networks
(with many layers) and thus the learning of rather complex tasks. See
\citet{Kaimang2015} for more details.

We use the ResNet-18 implementation provided by {\tt
  torchvision.models.resnet} and modify it in such a way that the
first convolutional layer of the model is expecting single-channel
(i.e., monochrome) images as input (instead of 3-channel RGB data) and
uses a $16\times16$ pixel convolutional kernel. Furthermore, the model
produces an output vector of length 33, representing the 33 subregions
in the image. As loss function, we choose the Binary Cross Entropy
with Logits Loss ({\tt pytorch.nn.BCEWithLogitsLoss}), which combines
calculating the binary cross entropy between the training features
(image data) and the training labels (presence of clouds) with a
Sigmoid layer for additional non-linear activation. As optimizer we use
Stochastic Gradient Descent with momentum \citep{Goodfellow2016}. 

\subsubsection{Training Procedure}


We train our model in single-batch mode 
using 70\% of the available training data (the remaining 30\% are used
as validation data sample) and start the optimization procress with a
learning rate of 0.025, which decreases at a rate of 0.3 every 5th
epoch for 100 epochs. 

The training is performed on a standard desktop computer that is
equipped with a NVIDIA GeForce GTX 1050 Ti graphics card, thus taking
advantage of Pytorch's GPU support. 

The results of the training are presented in
Section \ref{sec:results_resnet}.

\subsection{Feature-Based Approach (lightGBM)}

We also use a gradient-boosted tree-based model that works on a set
of features (see Section \ref{sec:data_features}) that we extract from
our images.  The term ``tree'' refers in this case to a decision tree,
which is a simple non-parametric machine learning model that can be
used for classification. A decision tree is a directional graph of
binary questions based on the provided feature space. In the case of a
classification task, each ``branch'' of the tree ends in a ``leaf'',
which defines a class assignment. See \citet{Russel2009} for a
discussion of decision trees.

A gradient-boosted tree-based model means the combination of a large
number of decision trees and an optimization rule that builds succint
trees to optimize the performance of the entire
ensemble. Gradient-boosted tree-based models are extremely flexible
and powerful in classification problems in pre-defined feature spaces.

We use the {\tt lightGBM} \citep{Ke2017} model implementation
in combination with the
{\tt sklearn.pipeline} infrastructure to train our model. We evaluate
sample scores using the cross entropy between model outputs and
training labels.

\subsubsection{Feature Definitions}
\label{sec:data_features}

We base the design of the feature space in which our tree-based model
will learn on human experience and mimick criteria that a human would
use in identifying clouds:

\begin{itemize}
\item source density: the presence of stars precludes the presence of
  thick clouds, which implies that subregions with a high density of
  sources most likely represent a clear patch of sky. We measure the
  number of sources per subregion with the Python module {\em SEP},
  which in turns makes use of {\em Source Extractor}
  \citep{Bertin1996}. Sources are identified based on a minimum number
  of pixel that are brighter than the background by a given threshold;
  tuning these parameters depends somewhat on the detector. The source
  count obtained using this method typically represents a lower limit,
  due to the usually rather limited imaging resolution of the all-sky
  camera. We derive the source density by dividing the number of
  sources per subregion by the total subregion pixel area.
\item background properties: depending on the illumination conditions
  (presence of the Moon or the Sun) clouds generally appear as dark or
  bright patches against the clear sky. We hence derive the average
  brightness, median brightness, and brightness standard deviation
  across each subregion.
\item time derivatives: clouds are rarely stationary; we take
  advantage of this property and form differences for each of the
  aforementioned features for each subregion. In order to be sensitive
  to both fast-moving and slow-moving clouds, we form subregion-based
  differences of each of the aforementioned properties between the
  property of the current image and that of images that were taken
  3~min ago and 15~min ago.
\end{itemize}

The proper interpretation of these features requires some additional
information, which we also add to the feature space:

\begin{itemize}
\item solar and lunar elevation, lunar phase: all three measures
  provide valuable information on the illumination circumstances and
  address the question whether the model should expect clouds to be
  darker or brighter than the clear sky.
\item subregion identifier: integer identifier of the current subregion,
  indicating its location on the local sky.
\end{itemize}

In total, the feature space considered encompasses 16 different
features for each subregion and image in the training data set.
In the training and prediction process we treat each subregion of each image
as an independent datapoint with an input vector of length 16 and a scalar
target value.

\subsubsection{Training and Hyperparameter Tuning}

We split the available labeled data sample into a validation sample,
containing 10\% of the examples, and another sample that is used in a
randomized cross-validation approach with 5 folds. We evaluate the
mean training and test scores in a uniform fashion over a wide range
in the most relevant model parameters: the number of estimators,
learning rate, maximum depth of each estimator, 
number of leaves per tree, and the minimum number of examples to form
a leaf. Regularization is achieved by sampling L1 and L2
regularization parameters $\alpha$ and $\lambda$ on a logarithmic
scale. The results of the training are presented in
Section \ref{sec:results_lightgbm}.

\section{Results}
\label{sec:results}

\subsection{ResNet}
\label{sec:results_resnet}

The performance of the ResNet model is somewhat sensitive to the
learning rate and momentum, but outcomes are very similar for learning
rates of the order of (1--3)\% and momentum values 0.7--0.9. However,
we do find significant variations between independent training runs
despite the use of manual random seeding, which we attribute to random
scheduling during the GPU acceleration and the relatively small
training sample size for this type of model. In the following, we
report on the results of the best of five independent training
runs.

\begin{figure}[t]
 \centering
 \plotone{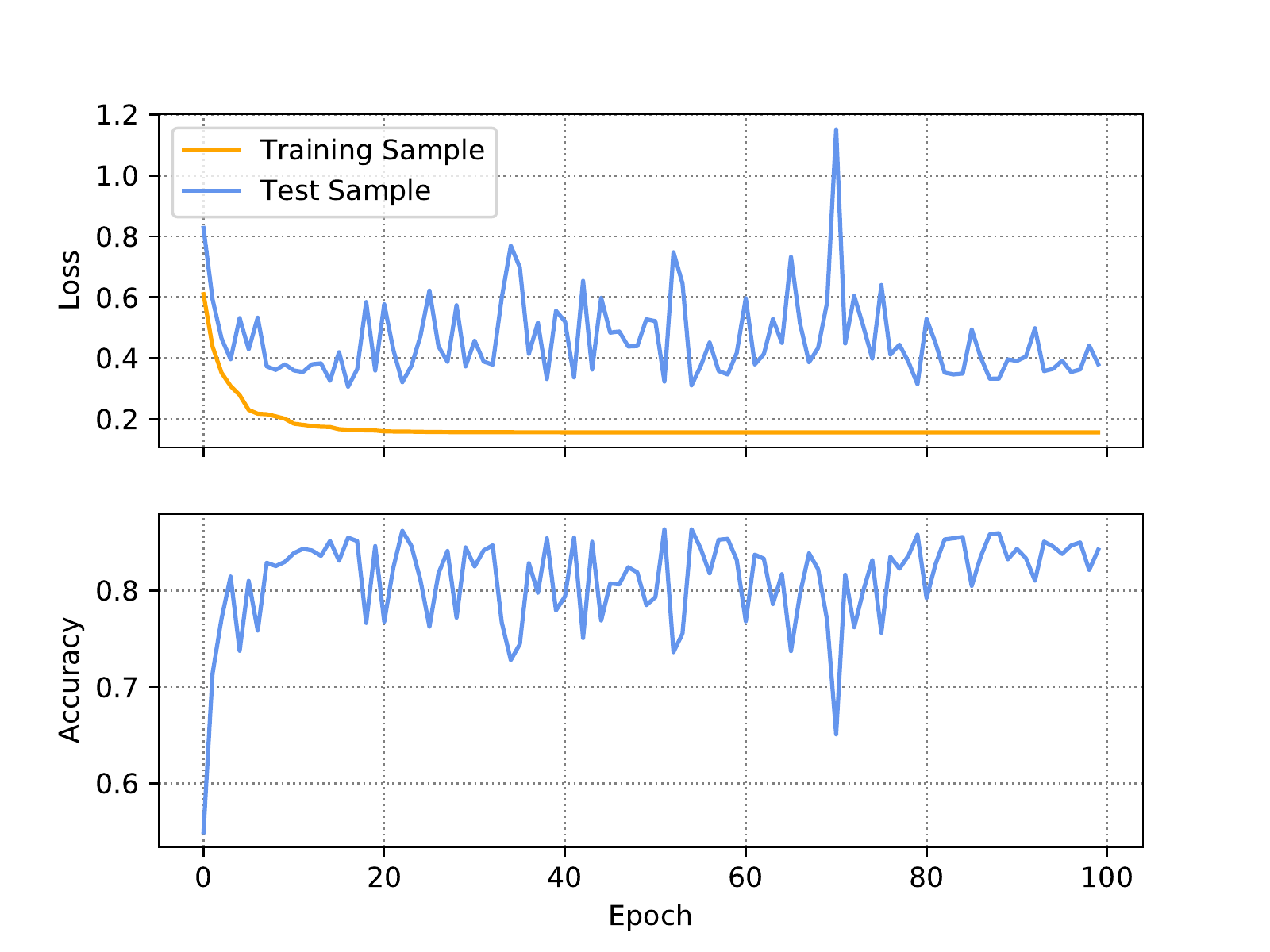}
 \caption{Loss (top) and test sample accuracy (bottom) of our ResNet
   model during 100 training epochs. After ${\sim}$20 epochs, the
   training sample loss is mostly stationary and the model barely
   improves. Given the low test sample loss and high test sample
   accuracy around 10 epochs, we recommend this number of epochs for
   training this model.}
\label{fig:resnet_training}
\end{figure}

Figure \ref{fig:resnet_training} shows that we find validation
sample accuracies of
the
order of ${\sim}$80\%, peaking around 87\% for individual training
epochs. After ${\sim}$20 epochs, the training sample loss becomes
mostly stationary, meaning that the model does not improve. The test
sample loss, however, is subject to significant variations, which we
attribute to the relatively small sample size. Training of the ResNet model
leads to rather high validation
sample accuracies of the order of 85\% after only ${\sim}$10 training epochs.
We adopt this accuracy and
number of epochs in our further analysis. We find f1-scores of the
order of 0.88. The f1-score is defined as the harmonic mean of precision and
recall and serves as a measure for the overall performance of a
binary classifier,
where 1 denotes a flawless classification and lower values denote flawed
classification results.

Training our ResNet adaptation for 100 epochs takes 6.9~hr, 10 epochs
of training takes accordingly 41~min.

\subsection{lightGBM}
\label{sec:results_lightgbm}

We adopt the following set of hyperparameters for our lightGBM model:
a maximum depth of each tree of 5, 500 estimators, a learning rate of
0.25, 30 leaves per tree, 100 examples required to form a leaf,
$\alpha=10$, and $\lambda=100$. This configuration leads to a training
sample accuracy of 96\% and a test sample accuracy of 95\%. The
accuracy on the validation sample, which was neither used in the
training of the model nor in the tuning of the hyperparameters, is
95\%, too. The f1-score on the validation sample is 0.94, underlining
the good performance of the trained model. 

The training of the entire training sample using the selected
hyperparameters and a 5-fold cross-validation takes 12~s on a standard
desktop computer.

\begin{figure}[t]
 \centering
 \plotone{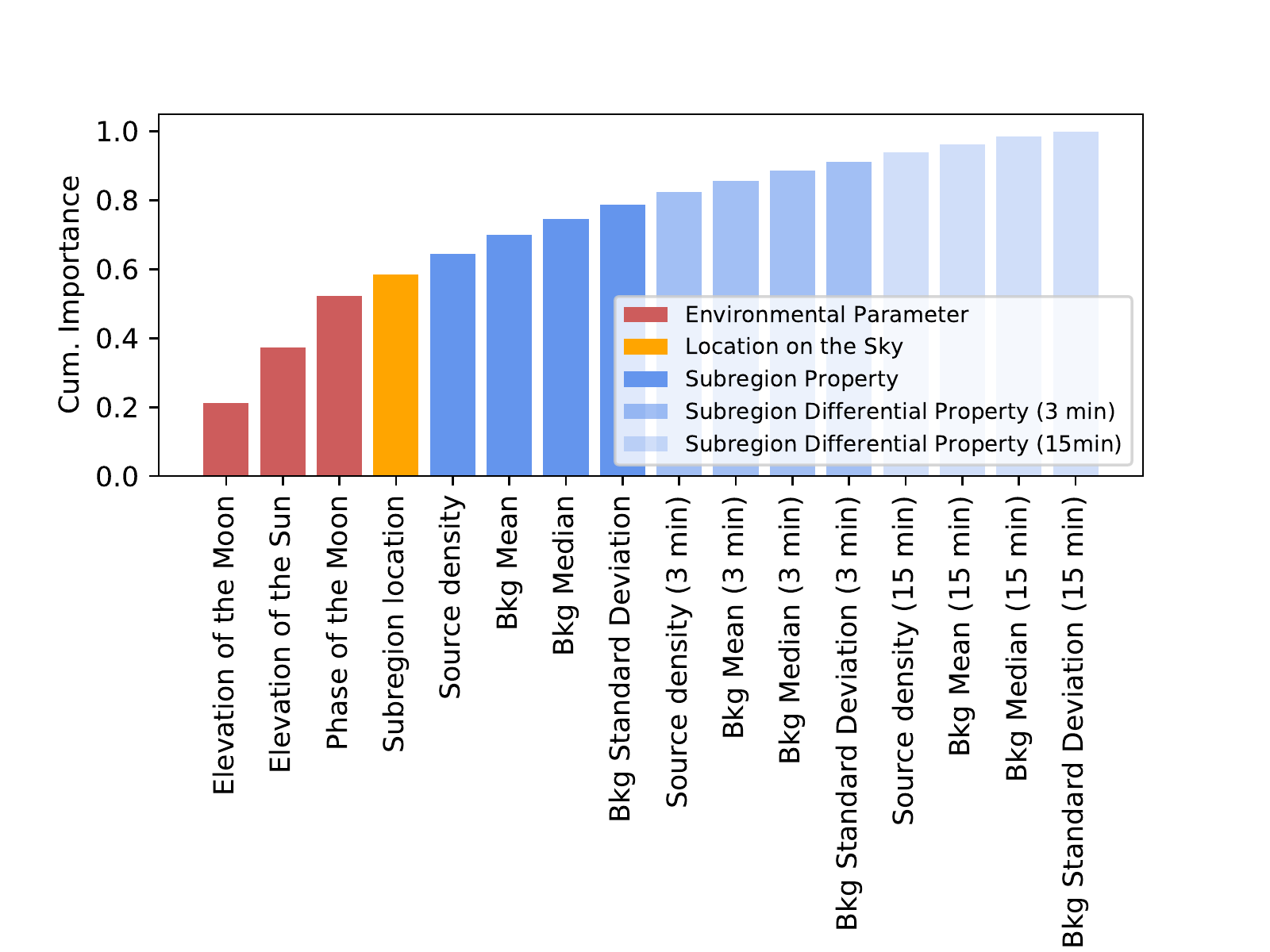}
 \caption{Cumulative feature importances extracted from the trained
   lightGBM model ordered by magnitude. Environmental parameters
   affecting the sky brightness play a major role in the
   identification of clouds, followed by the location in the
   sky. Subregion properties and their differentials play a
   decreasingly important role with time. }
\label{fig:lightgbm_featureimp}
\end{figure}

Figure \ref{fig:lightgbm_featureimp} shows the feature importances
extracted from the final trained model. The feature importance used
here is defined as the number of times a feature is used in this model
throughout all individual decision trees. The comparison shows that
environmental parameters that affect sky brightness are extremely
important, followed by the subregion location. Actual subregion
properties and their time differentials follow, the latter of which
only have a small -- but not negligible -- impact on the model
results.

\section{Discussion}
\label{sec:discussion}

\subsection{Model Performance}

We find the feature-based approach using lightGBM to be both
significantly faster and more accurate than our image-based approach
using the ResNet adaptation. Only our lightGBM approach meets our
requirement of properly identifying 95\% of subregions that contain
clouds. This discrepancy can be explained with the fact that our
feature-based approach takes advantage of pre-defined features guided by human
experience which the ResNet model has to learn by itself.

While the feature-less approach using our ResNet model shows some
promise, this model most likely requires much more training data to
reach a similar accuracy as our lightGBM model. However, the need of
more training data, which has to be labeled manually, in combination
with the much higher computational requirements, make this deep
learning approach much less attractive.

\subsubsection{Model Accuracy and Confusion Matrix}
\label{sec:discussion_accuracy}

The cloud detection probability for a single subregion is ${\sim}85$\%
using ResNet and ${\sim}$95\% using lightGBM. Since clouds typically
cover more than one subregion, the probability that any subregion in a
set of $N$ subregions that actually include clouds increases
exponentially with $N$. In the same way, the probability to miss
clouds decreases. For example, the probability to miss the detection
of clouds in three different subregions with the lightGBM classifier
is ${\sim}0.05^3=10^{-4}$. Hence, the confidence in detecting the
presence of clouds anywhere on the sky is much higher than the
probability to detect them in a single subregion, supporting the usefulness
of this machine-learning approach.

\begin{figure}[t]
 \centering
 \plotone{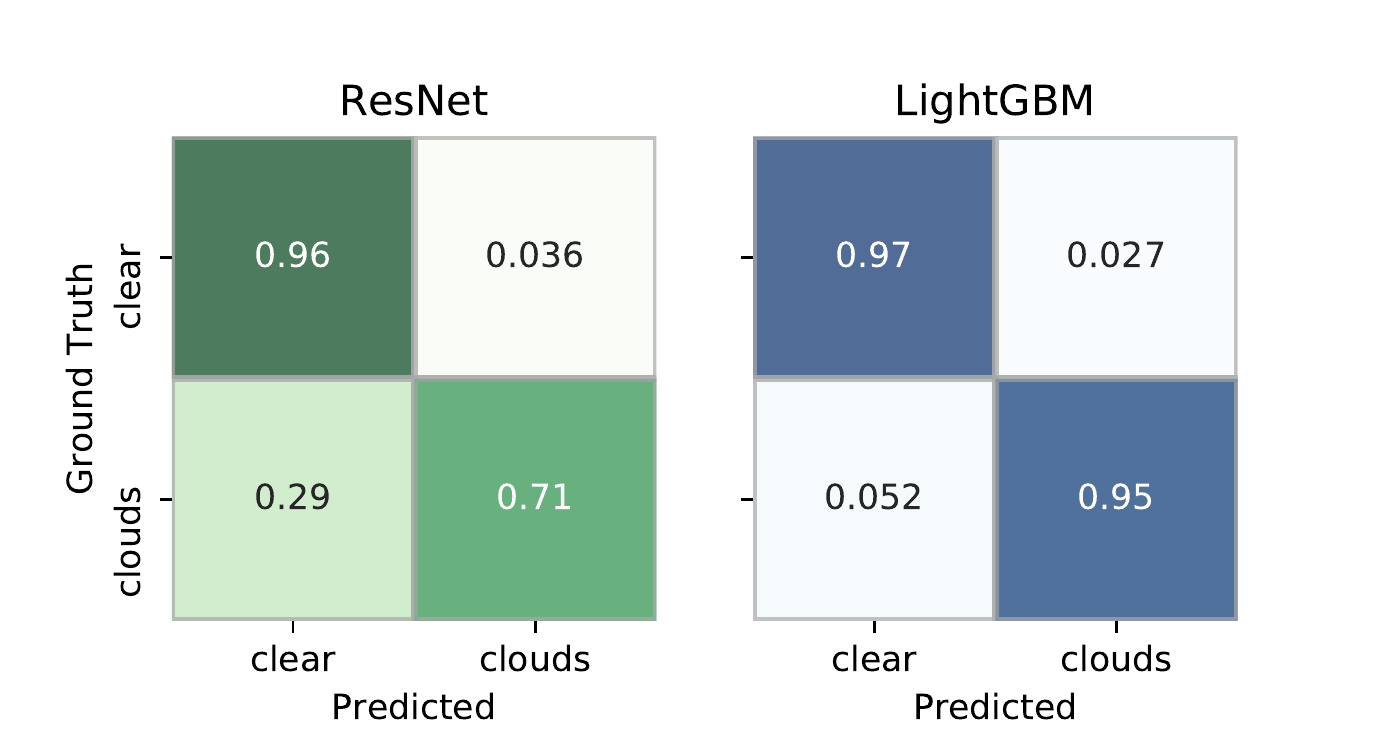}
 \caption{Confusion matrices for the methods considered in this work. }
\label{fig:confusion_matrix}
\end{figure}

We further investigate the performance of our models using a confusion
matrix, which not only provides information on the overall classification
accuracy, but also additional information on the rate of false positive and
false negative classifications. Here, a false positive classification means a
subregion that has been predicted to contain clouds, although this is
not
the case. A false negative classification refers to a subregion that contains
clouds which are not identified by the classifier. The confusion matrices for
both methods used here are shown in Figure \ref{fig:confusion_matrix}.

As Figure \ref{fig:confusion_matrix} shows, the false negative and false
positive rates using the lightGBM classifier are rather small at
5.2\% and 2.7\%, respectively. We point out that the false negative rate is
roughly a factor 2 higher than the false positive rate, which might be
slightly affected by the class imbalance inherent to the training data sample
(see Section \ref{sec:data_training}), but is mostly likely due to the
classifier's inability to identify non-opaque clouds that were labeled in the
training data set. This effect, as well as additional shortcomings
potentially related to the insufficient size of the training sample (see
Section \ref{sec:discussion_training_data_needed}), is much more
obvious in the  results of
the ResNet classifier, which achieves a false negative rate of 29\% and a false
positive rate of 3.6\%, underlining the insufficient performance of this
classifier. The comparison of these numbers support the suitability of the lightGBM
approach for this task, which is able to identify clouds with high confidence.

We note that for both model approaches the rate of misclassifications
(false positives and false negatives) is highest for subregions close
to the horizon. This is most likely due to confusion with layers of
haze or other near-surface effects, as well as human subjectivity
introduced in the training sample (see Section
\ref{sec:cloud_definition}). This issue is most likely to be resolved
with more consistent manual labeling of a larger training sample.

\subsection{Training Data}

The performance of each model depends highly on the amount and quality
of the available training data sample.

\subsubsection{How much training data is needed?}
\label{sec:discussion_training_data_needed}

We investigate the impact of the training data sample size on the
model performance by training the same models on random subsamples of
the original training data sample.  We use the same sets of
hyperparameters presented in Section \ref{sec:results}. Results for
different subsample sizes are listed in Table \ref{tab:trainingsize}.

\begin{deluxetable*}{c|cc|cc}
\tablecaption{Model Performance as a Function of Training Sample Size.\label{tab:trainingsize}}
\tablewidth{0pt}
\tablehead{
\colhead{Training Sample Size} & \multicolumn2c{ResNet} & \multicolumn2c{lightGBM} \\
\colhead{(N$_{\mathrm{images}}$)} & \colhead{Accuracy} & \colhead{F1} & \colhead{Accuracy} & \colhead{F1}
}
\startdata
1975 & 0.843 & 0.877 & 0.946 & 0.937 \\
1000 & 0.816 & 0.847 & 0.945 & 0.939 \\
500  & 0.762 & 0.795 & 0.932 & 0.919 \\
100  & 0.552 & 0.591 & 0.915 & 0.910 \\
\enddata
\end{deluxetable*}

In the case of our ResNet approach, a steady rise of both accuracy and
F1-score can be observed through all training sample sizes used in
this analysis. The fact that neither metric plateaus indicates that
the trainig sample size required to max out the performance of the
ResNet model has not been reached and that this model will benefit
from additional training data. We fit a power-law function of the form
$f(x; a, b, c)=a-b\cdot10^{c}$ to the ResNet accuracy values in Table
\ref{tab:trainingsize} and find a saturation accuracy of only 92\%.
Furthermore, we find through extrapolation that our ResNet
approach requires of the order of 20,000 training samples to achieve
an accuracy of 90\%. We acknowledge that this extrapolation may not be
highly accurate, but it certainly provides reasonable estimates of the
orders of
magnitudes for both the maximum accuracy that can be expected and the
training sample size. Based on these estimates, we conclude that our
ResNet approach in this form is extremely expensive compared to our lightGBM
model.

We find lightGBM performances that are comparable to those reported in
Section \ref{sec:results_lightgbm} for training sample sizes of the
order of 1000 image examples. Even in the case of only 100 image
examples, an accuracy above 90\% can be reached. This result implies
that the lightGBM approach is useful even if only a small training
sample is available. We furthermore conclude that more then 1000
training examples will not significantly improve the performance of
this model.

\subsubsection{Training Data Quality and Cloud Definition}
\label{sec:cloud_definition}

The training data sample should contain as little noise as possible in
order to maximize the model performance. However, it is not always
possible to provide unambiguous training data in the case of cloud
identification, as it is a highly subjective process even for a human.

Clouds have many different ways to manifest in all-sky camera images,
making it nearly impossible to come up with a clear definition what
counts as a cloud and what not. Does thin cirrus count as a cloud? Do
you consider a subregion to contain a cloud if it occupies less than
10\% of that subregion's area? Does haze on the horizon count as a
cloud? There is no definitive answer to these questions, adding a
significant amount of noise to the training data sample.

The sensitivity of any machine learning effort to cirrus and other not fully
opaque clouds depends highly on the training data provided. Based on the goal
set for this work -- the implementation of a cloud warning system -- we
chose a rather conservative cloud definition in the sense that we expect a cloud
to be fully
opaque. While this definition is less prone to human subjectivity (leading
to a more homogeneous training sample), it clearly creates a bias in the
quantification of sky quality (e.g., cirrus is likely to
be not detected as a cloud), which becomes apparent in the false negative
rates of both classifiers (see Section \ref{sec:discussion_accuracy}). A less
conservative
cloud
definition can lead
to a better detectability of cirrus, but might be susceptible to other
phenomena like air glow and terrestrial light sources. Additional processing
of the input data might be necessary to distinguish the latter two effects
from cirrus, e.g., by exploiting their static nature in the night sky. We
leave such investigations for the future.

Both of our models utilize regularization mechanisms to be able to
generalize the training data and to cope with noise in the training
data. However, this also means that the subjective uncertainty of
humans is propagated into the models: if the presence of clouds in a
given subregion is vague to a human, it will also be vague to the
model trained on data labeled by a human. Any labeling efforts by
humans should thus be as consistent as possible.

We hence believe that the performance any model can achieve is mainly
limited by the quality of the training data sample.

\section{Conclusions}

We find that the identification of clouds in all-sky camera data is a
solvable task for machine learning models. We use two different
approaches for this task: a ResNet model that works with image data
and achieves an accuracy of ${\sim}85$\%, and a lightGBM model that
uses features extracted from the images and achieves an accuracy of
${\sim}$95\%. While we find false negative and false positive rates of
only a few percent for the lightGBM model, the ResNet model has a false
positive rate of ${\sim}$3\% and a false negative rate of 29\%.
These estimates are based on a training data sample
containing 1975 images. While we expect a slightly better performance
for the ResNet model wih a larger sample size, the lightGBM model seems
to require only 1000 training samples to obtain its full performance.

In conclusion, we recommend the use of feature extraction in
combination with a simple classification model, like lightGBM, as it
provides superior performance in terms of accuracy and runtime.

The methods presented here are tailored to the detection of opaque clouds,
i.e., for the
protection of observatory equipment from weather. This is mostly achieved
through the cloud definition that is applied in the labeling of the training
data set. Additional steps will have to be
taken to extend the usability of this methods for automated sky quality
quantification that is also able to detect even thin cirrus or photometric
conditions.

\acknowledgments

The author would like to thank Ryan J. Kelly and the NAU/NASA Arizona
Space Grant program for enabling a case study for this project.

\software{astropy \citep{astropy2013, astropy2018},
          Django (\url{https://djangoproject.com}),
          lightGBM (\url{https://github.com/microsoft/LightGBM}),
          matplotlib \citep{Hunter2007},
          numpy \citep{Oliphant2006},
          pandas \citep{McKinney2010},
          pytorch \citep{Paszke2019},
          scipy \citep{Virtanen2019},
          seaborn \citep{Waskom2020},
          SEP (\url{https://github.com/kbarbary/sep}),
          SExtractor \citep{Bertin1996},
          scikit-image \citep{vanderWalt2014},
          scikit-learn \citep{Pedegrosa2011}
          }

\appendix

\section{Implementation and Resources}
\label{sec:implementation}

The code built for this work is publicly available under a 3-clause BSD-style
license at \url{https://github.com/mommermi/cloudynight} \citep{Mommert2020}. The {\tt cloudynight}
repository contains (1) a Python module for
data handling and preparation, feature extraction, model training and
prediction, (2) a Python Django web server application for database
management, data visualization, and manual labeling, (3) example
data used in this work, and (4) a number of scripts for testing the
functionality on the example data.

Please note that this code is not intended for
plug-and-play. Instead, it is tailored to the example data used in this work.
However, with the descriptions in this appendix and comments provided as part of
the code, it should be easy to modify the code of the module and the web
application to run them on other data sets.

The example data provided are sufficient to test the different parts of the
provided code. However, especially the image data are not sufficient to train
a meaningful model. Also note that due to the small number of images included
in the example image set, it is impossible to derive time-dependent features
from images -- these features are hence set to zero
in this case.

\subsection{Design, Premises, and Setup}

The {\tt cloudynight} module consists of three classes: {\tt AllskyImage},
which handles individual allsky images, {\tt AllskyCamera}, which handles
image management and provides wrapper methods for image sets, and {\tt
LightGBMModel}, which handles the lightGBM model used in this work. Note that
the ResNet model implementation is provided as an example script, but not as
part of {\tt cloudynight}. A large fraction of the module's setting are
configurable using parameters defined by {\tt cloudynight.conf} (in the file
{\tt \_\_init\_\_.py}).

Camera images are expected in FITS format. Each image should have a header
containing at least the date and time of observation (header keyword {\tt
DATE-OBS}). Camera images are also expected to be sorted by night and to reside on a
remote machine. The latest data for a given night can be
downloaded from that remote machine with
{\tt AllskyCamera.download\_latest\_data}, which
uses an {\tt rsync} call to only copy new data; this mechanism requires the
exchange of ssh keys between both machines. Downloaded images for a given
night can
be processed (prepared and feature-extracted) and results can be uploaded to
the database with {\tt AllskyCamera.process\_and\_upload}.

The database has to be setup as part of the Django web application. The use
of array fields in the database requires a {\tt PostgreSQL} database type.
The setup of this web application is detailed in the repository
documentation. The web application provides a RESTful API that can be used
to access and modify the database from any other machine on the network,
e.g., to run a trained model on a given image to detect clouds.

The database contains 3 tables:

\begin{itemize}
\item {\tt Subregion}: contains outlines of the individual subregions for
      proper display in the web application;
\item {\tt Unlabeled}: contains image features for images that have not been
      labeled manually;
\item {\tt Labeled}: contains image features and cloud labels for images that
      have been labeled manually; this is the training data set.
\end{itemize}

Before using the web application, an image mask has to be created using the
script {\tt generate\_mask.py} and the subregion outlines have to be uploaded
to the database using the script {\tt subregions.py}.

\subsection{Training}

After the proper setup of the module and the web application, training data
have be generated. For this purpose, data from a large number of nights
should be downloaded, processed, and uploaded to the {\tt Unlabeled} table of
the database using {\tt AllskyCamera.process\_and\_upload\_data}. The
training
task ({\tt label/}) of the web application can
now be used to manually label
subregions that contain clouds; results are automatically saved to the {\tt
Labeled} table of the database.

\subsection{Model Fitting}

If enough training data are available, a model can be fitted The script
{\tt model\_lightgbm.py} can be used as a template for this task. Model
performances
for different model parameters can be explored with
{\tt LightGBMModel.train\_randomizedsearchcv}. Once a model solution has been
found, the model has to be saved as a file using
{\tt LightGBMModel.write\_model} so that the trained model can be utilized by the web
application.

\subsection{Cloud Detection}

The presence of clouds can be predicted for the latest camera using the web
application task {\tt predictLatestUnlabeled/}. This task
returns a json
object that contains an array in which each element represents one subregion;
the value of the element will be 1 in case a cloud has been detected and 0
otherwise. This information can be utilized by the user in any possible way.

A different way to predict the presence of clouds would be to download
data using the RESTful API and then run a model in a Python script similar to
{\tt model\_lightgbm.py}.

For automated cloud prediction on the latest camera image, the user has to
setup a cron job that runs a script to download the latest image data for
the current night and detects clouds in this image using
{\tt predictLatestUnlabeled/}.

\end{document}